\documentclass[conference]{IEEEtran}
\IEEEoverridecommandlockouts

\pdfoutput=1
\usepackage{algorithm}
\usepackage[noend]{algpseudocode}
\usepackage{cite}
\usepackage{graphicx}
\usepackage{lipsum}

\usepackage{xspace}

\newcommand{\BlockSim}{BlockSim\xspace}

\begin{document}

\title{Data-Driven Model-Based Analysis of the Ethereum Verifier's Dilemma\thanks{\textbf{This is the author’s version of the work. The final version will be published in the proceedings of the 50th IEEE/IFIP International Conference on Dependable Systems and Networks (DSN 2020).}}\thanks{
This study was financed in part by the Coordena\c{c}\~ao de Aperfei\c{c}oamento de Pessoal de N\'ivel Superior - Brasil (CAPES) - Finance Code 001. Also, this paper was funded by IFRS and was achieved in cooperation with HP Brasil using incentives of Brazilian Informatics Law (Law n 8.248 of 1991).}
}


\author{\IEEEauthorblockN{Maher Alharby\IEEEauthorrefmark{1}\IEEEauthorrefmark{2},
Roben Castagna Lunardi\IEEEauthorrefmark{3}\IEEEauthorrefmark{4}, Amjad Aldweesh\IEEEauthorrefmark{1}
Aad van Moorsel\IEEEauthorrefmark{1}}
\IEEEauthorblockA{\IEEEauthorrefmark{1}School of Computing, Newcastle University, Newcastle upon Tyne, UK}
\IEEEauthorblockA{\IEEEauthorrefmark{2}College of Computer Science and Engineering, Taibah University, Medina, KSA}
\IEEEauthorblockA{\IEEEauthorrefmark{3}Pontificial Catholic University of Rio Grande do Sul (PUCRS), Brazil}
\IEEEauthorblockA{\IEEEauthorrefmark{4}Federal Institute of Rio Grande do Sul (IFRS), Brazil}
\IEEEauthorblockA{Email: \{m.w.r.alharby2, r.castagna-lunardi2, a.y.a.aldweesh2, aad.vanmoorsel\}@ncl.ac.uk}}


\maketitle

\begin{abstract}
In proof-of-work based blockchains such as Ethereum, verification of blocks is an integral part of establishing consensus across nodes. However, in Ethereum, miners do not receive a reward for verifying. This implies that miners face the Verifier's Dilemma: use resources for verification, or use them for the more lucrative mining of new blocks? We provide an extensive analysis of the Verifier's Dilemma, using a data-driven model-based approach that combines closed-form expressions, machine learning techniques and discrete-event simulation. We collect data from over 300,000 smart contracts and experimentally obtain their CPU execution times. Gaussian Mixture Models and Random Forest Regression transform the data into distributions and inputs suitable for the simulator. We show that, indeed, it is often economically rational not to verify.  We consider two approaches to mitigate the implications of the Verifier's Dilemma, namely parallelization and active insertion of invalid blocks, both will be shown to be effective.    
\end{abstract}


\section{Introduction}
Blockchains depend on miners to operate the chain correctly and to jointly guarantee consistency and correctness of the blockchain data and the executed transactions.  In public, permissionless, blockchains collaboration of miners is based on incentive mechanisms that provide miners with a certain amount of cryptocurrency for their efforts. It is clearly important to award fees in such a manner that correct and desired behaviour is encouraged. Well-balanced incentives, together with the miner's interest to keep the system running well, should guarantee the correct behaviour of the blockchain.  

Within Ethereum, there is one interesting aspect of the consensus algorithm that is not incentivised directly, namely the verification of transactions and blocks. This leads to an interesting dilemma: should miners verify transactions within blocks if they do not receive a specific fee for it?  If all blocks are valid, the verification would not have been necessary and the time spent on verifying could have been used to mine new blocks (which are rewarded by a fee).  This Verifier's Dilemma is well recognized, e.g., \cite{luu2015demystifying,teutsch2017scalable}, but has not been systematically analysed. In this paper we conduct that systematic analysis of the Verifier's Dilemma in Ethereum.  

The analysis is involved, and combines a number of analysis techniques to establish the fees miners would collect under different decisions about participation in the verification.  We pursue a model-based approach, so that we are able to analyse a range of possible scenarios.  It is not practical or even possible to obtain insights about the Verifier's Dilemma solely based on observations of the actual Ethereum system.  A model-based approach, correctly intertwined with data-driven parameterization, is the only reasonable approach. 

We combine the following techniques.  At the core of the analysis is the publicly available Ethereum simulator \BlockSim. This simulator is general purpose and extensible by design \cite{alharby2019blocksim}. We extended it with functionality necessary for the analysis of the Verifier's Dilemma under various scenarios.  Secondly, to run realistic simulation studies, we conduct an extensive data collection exercise for Ethereum smart contracts. This data aims to feed into the simulator realistic system parameters and realistic characteristics of smart contract based transactions. The data collection includes 324 thousand Ethereum smart contracts. Thirdly, we applied statistic/machine-learning techniques to the data to make it suitable as input to the simulator.  In particular, we use Gaussian Mixture Models to fit distributions to the data, for instance the distribution of Used Gas per smart contract (for Used Gas, see Section \ref{s:background}). And we use Random Forest Regression to predict the CPU time needed to execute smart contracts, given the Used Gas. The resulting distributions are used to parameterize the simulations.  As a fourth and final element in our study, we obtain a number of closed-form results for base scenarios.  In these base scenarios no invalid blocks are present, and under that assumption we are able to derive expressions for the rewards miners receive if they do or do not verify blocks. 

The conclusion of the above analysis is that under certain conditions it pays off for miners not to verify. Obviously, not verifying blocks put the correct functioning of the blockchain at risk, since the consensus approach in Ethereum assumes verification to take place.  To mitigate this risk, we consider two approaches.  First, we consider parallel verification (as proposed in \cite{dickerson2017adding}), to decrease the time it takes to verify blocks and therefore decrease the time verifying miners would have to spend before they can mine a new block.  Secondly, we will consider the idea of injecting invalid blocks on purpose, to penalize miners that do not verify.  The reasoning behind that approach was identified in \cite{teutsch2017scalable}.  By injecting invalid blocks, a non-verifying miner would more often pass on chains with invalid blocks that will be rejected by other miners, which in turn would imply that the non-verifying miner does not receive the block award. 

To summarize, the results of our analysis are as follows. It is clear that there are many scenarios in which miners would benefit from not verifying blocks.  This is especially true if (1) all or almost all blocks are in fact valid, and (2) if the block limit is large enough.  For Ethereum, currently the impact is small but the Verifier's Dilemma will become more important when the block limit increases, as is anticipated \cite{luu2015demystifying}. As mitigation approaches, both parallel verification and injecting invalid blocks improve the situation.  That is, both make it less lucrative for miners to avoid verifying. 

This paper is structured as follows.  Section \ref{s:background} and Section \ref{s:dilemma} introduce the problem space, providing background and an explanation of the Verifier's Dilemma.  Section \ref{ss:model} provides the closed-form expressions for the gain in rewards non-verifying members could get for the base model, i.e., the case that all transactions are valid. Section \ref{s:mitigation} introduces the two mitigation approaches, namely parallelization and injecting invalid transactions.  The data collection exercise for over 300,000 smart contracts in Ethereum is discussed in Section \ref{s:bench}.  Section \ref{ss:Fitting} also discusses the machine learning approaches of Gaussian Mixture Models and Random Forest Regression used to establish distributions that serve as input for the simulator.  We note that more detailed results derived with the machine learning techniques can be found in Appendix \ref{s:append}. Section \ref{s:evaluation} describes how we used the publicly available simulator and enhanced it to suit our study.  The results of the simulation study are provided in Section \ref{s:results}, for various scenarios as well as for the two mitigation strategies and the insights gained and conclusions drawn are further discussed in Section \ref{s:discussion}.  Finally, related work is discussed in Section \ref{s:related} and Section \ref{s:conclusion} provides the conclusion.

\section{Background}
\label{s:background}

\subsection{Ethereum Blockchain}
\label{ss:eth}

A blockchain is a distributed ledger of linked blocks, where each block contains a number of transactions. In the blockchain network, a set of nodes, known as miners, are responsible for maintaining the ledger by continuously appending new blocks. To generate and append a new block to the blockchain, miners first select and execute a number of pending transactions from the network and then include them in the block by executing a mechanism such as Proof of Work (PoW). The generated block will then be propagated to other nodes in the network. Upon receiving the newly generated block, every node is expected to verify the block before adding it to its blockchain's copy. Verifying a block requires checking the correctness of the block construction (e.g., PoW verification) and executing all transactions included in the block to check the outcome. This is to verify and validate miners' work, and is called the verification process.

The first generation of blockchain systems (e.g., Bitcoin \cite{nakamoto2008bitcoin}) provided predominantly cryptocurrencies in support of financial interactions. Blockchain-based smart contract systems (e.g., Ethereum \cite{white,wood2019}) emerged to support more complex distributed applications through smart contracts. A smart contract is essentially a computer program that can be attached to the blockchain. It contains contractual clauses and it is enforced by the consensus algorithm. Currently, Ethereum is the most popular smart contract platform \cite{hegedus2018popular}, offering a Turing-complete programming language for writing contracts. 

In Ethereum, there are two types of accounts (users), namely, externally owned accounts and contract accounts. Contract accounts contain smart contract code and some storage space to support the execution of smart contracts.  Accounts interact with each other through transactions. In Ethereum, there are two types of transactions, namely, transfer and contract transactions. The former is to move Ether (the Ethereum cryptocurrency) between accounts, while the latter is to either publish a new smart contract to the blockchain or to invoke an existing one.

To publish a new contract, a contract-creation transaction containing the creation bytecode for the contract is submitted to the blockchain. Once the transaction is executed, the contract will be deployed and a unique address will be assigned to it. To invoke that contract, a contract-execution transaction attaching appropriate input data is sent to the contract's address. The input data is the contract's function to be executed and its arguments. 

All nodes in the Ethereum blockchain run the Ethereum Virtual Machine (EVM). The EVM is responsible for executing contract transactions \cite{wood2019}. It executes the smart contract instructions, known as opcodes. Transactions are currently executed and verified on the EVM sequentially.

\subsection{Ethereum Incentive Model}
\label{ss:gas}

Ethereum has a built-in incentive model to reward miners for maintaining and expanding the blockchain ledger. There are three types of rewards, namely, block reward, block's transactions fees and uncle rewards. The block reward is a fixed amount of Ether (currently, 2 Ether) for each new block. The block's transactions fees are the fees associated with all transactions included in the block. The uncle rewards are a fraction of rewards for generating and including a new block that turns into a stale (orphan) block \cite{wood2019}.

Ethereum uses the Gas mechanism to calculate the fee for smart contract transactions. Each opcode of a smart contract uses a predefined amount of gas, as specified in \cite{wood2019}. The EVM tallies the amount of Used Gas and charges the submitter of the transaction based on the Used Gas. To avoid non-terminating transactions the submitter specifies a Gas Limit, and the EVM stops processing if that limit is reached (in which case Used Gas $=$ Gas Limit). The submitter also specifies a Gas Price (expressed in Ether) and the miner then charges the submitter the following transaction fee: Used Gas $\times$ Gas Price. The more opcodes the transaction requires, the more CPU effort from the miner, but also the higher the received reward. 

It is crucial to ensure that the Ethereum incentive model is fair in order to keep miners well-motivated to participate and to maintain the blockchain ledger honestly \cite{aldweesh18} \cite{alharby18survey}. 
Otherwise, miners may prefer to deviate from the desired behaviour. One issue of Ethereum and some other blockchains is that there are {\em no miner incentives for verifying} the recipient blocks. As a result, miners might be encouraged to avoid the verification process, especially if it tends to be computationally intensive, in favour of maximizing their revenues, as we will discuss in the following section.

\section{Verifier's Dilemma in Ethereum}
\label{s:dilemma}
We first present and discuss in Section \ref{ss:verifier} the Verifier's Dilemma in general terms and then we derive in Section \ref{ss:model} closed-form expression for the rewards received by non-verifying miners.  These closed-form expressions hold for scenarios in which all blocks are valid, which we will call the {\em base model}. 

\subsection{Problem Description}
\label{ss:verifier}
Luu et al. \cite{luu2015demystifying} pointed out that verification of blocks consumes computation resources and time, and thus, delays miners in the race of mining the next block. Not only it delays miners, but also it does not provide incentives (a free task) to miners. This is especially true in smart contract based blockchains since verification of smart contracts involves repeating the execution of the smart contract to check the outputs. \cite{luu2015demystifying} also points out that these concerns exacerbate when the block limit increases since that increases the number of transactions to verify.  

As a result, miners might consider skipping the verification process. Skipping verification allows the miners to turn to the profitable mining of new blocks. The risk of skipping verification is that the miner adds its newly mined blocks to a blockchain that contains invalid blocks.  If other miners verify these blocks they will disregard these and the new block and the non-verifying miner will not receive a reward for its new block.  The miner needs to decide the following: should I support the blockchain honestly and verify all blocks, possibly at the cost of personal rewards, or shall I skip verifying blocks and instead spent the time on the lucrative mining of new blocks, thus increasing personal rewards? \cite{luu2015demystifying} calls this the {\em Verifier's Dilemma}. 

The Verifier's Dilemma has received some attention, we refer to the discussion on related work in Section \ref{s:related}.  However, there has not been a rigorous analysis of the dilemma using probabilistic modelling techniques such as in this paper.  

\subsection{Ethereum Base Model}
\label{ss:model}
In Ethereum, miners are expected to verify received blocks by executing the block's transactions in sequence. In this section, we use closed-form solutions to investigate the Verifier's Dilemma in Ethereum and its impact on the fee received by miners. We consider current and likely future configurations of Ethereum in terms of block limit and block interval time. 

There are a few model assumptions that will hold throughout the paper. We assume that miners fill each block by executing as many transactions as they can in order to maximise their revenue. In the real system, miners can generate full, non-full, or even empty blocks, but this is not critical for the analysis of the Verifier's Dilemma.  If needed this can be added to the model. All transactions in the network are assumed to be contract-based transactions, thus ignoring the additional financial transactions that may take place.  Such financial transactions take less time to verify and therefore do not impact the Verifier's Dilemma as much, but these can of course easily be added.  We also ignore the time it takes to check the hash outcome of the PoW, since that check is almost immediate.  
Finally, we do not explicitly consider block propagation delay between nodes since this does not affect the issue of the Verifiier's Dilemma, 

We will now derive closed form results for scenarios in which all miners are honest when executing transactions. That is, all transactions included in a block are valid.  Assume that all miners use the same hardware/software architectures, and thus, the CPU time required to execute transactions is the same for all miners. The verification time $T\textsubscript{v}$ is the CPU time required to execute and verify all transactions embedded in a block. It is worth noting that a miner only verifies blocks that are generated by other miners, not the ones it generates itself. Thus, the average block's verification time decreases with the increase of the miner's hash power $\alpha$, since then there are less blocks generated by others. For instance, a miner with $\alpha$ = 0.30 of total network hash power is expected to verify 70\% of the total generated blocks, which means that it spends on average $(1-\alpha) * T\textsubscript{v}$ for verification per block. 

We propose the following closed-form solutions to estimate the fraction of fee received by both verifying and non-verifying miners as a function of block limit, block interval time and fraction of hash power. We estimate the  slow down $\delta$  of performing the sequential verification process by considering the average time required to verify the received blocks $T\textsubscript{v}$ and the sum of the hash powers of all verifying miners $\alpha V$ as follows.

\begin{equation}
    \delta = (1 - \alpha V) * T\textsubscript{v}
\label{eq:delta}
\end{equation}
Where $1 - \alpha V$ represents the fraction of hash power of non-verifying miners. The probability of finding the next block by a verifying miner is then reduced from $\frac{\alpha v}{T\textsubscript{b}}$ to $\frac{\alpha v}{T\textsubscript{b} + \delta}$, where $T\textsubscript{b}$ is the block interval time. That means the fraction of the expected blocks and rewards ($Rv$) for a verifying miner is reduced from $\alpha v$ to
\begin{equation}
    Rv = \frac{\alpha v}{T\textsubscript{b} + \delta}  * T\textsubscript{b}
\label{eq:Rv}
\end{equation}

The fraction of the expected blocks and rewards ($Rs$) for a non-verifying miner is increased from $\alpha s$ to
\begin{equation}
    Rs = \alpha s + \frac{\alpha s     (\alpha V - RV)}{\alpha S} 
\end{equation}
Where $\alpha S$ is the sum of the hash powers of all non-verifying miners, and $RV$ is the fraction of blocks and rewards gained by all verifying miners.

Results from this closed-form solution will be discussed in Section \ref{s:results}, but to illustrate the implications, assume 10 miners, each controlling $\alpha$ = 0.1 of the total network hash power. Assume there is only one miner among those miners who does not verify. 
Assume $T\textsubscript{v}=3.18$ and $T\textsubscript{b}=12$ seconds. We calculate the slow down of performing the verification as $\delta = 0.318$. The fraction of blocks generated by the nine verifying miners is reduced from 0.9 to 0.878. Thus, the non-verifying miner would gain 0.022 more blocks and rewards ($\approx$ 22\% more than its invested $\alpha$). In other words, the fraction of reward obtained by the non-verifying miner is increased from 0.1 to 0.122.

\section{Mitigation Solutions to the Verifier's Dilemma}
\label{s:mitigation}
We discuss two mitigation solutions for the Verifier's Dilemma, namely parallel verification of transactions and intentional production of invalid blocks.

\subsection{Mitigation 1: Parallel Verification}
\label{ss:par_model}
Parallel verification was proposed by \cite{dickerson2017adding} to speed up the verification process, thereby minimizing the lost time to miners.  By speeding up the time it takes to verify transactions, a miner would loose less time. Transactions that are not in conflict (read/write conflicts) with other transactions in the same block can be verified in parallel. The remaining conflicting transactions must still be verified in sequence.

To implement parallel verification in a real system, the Ethereum Virtual Machine needs to support it, using multi-threading. Miners then attach an execution schedule to their proposed blocks. The schedule details which transactions can be processed in parallel (no read/write conflicts) and which must be executed in sequence. We assume miners provide a correct schedule and are well motivated to include the schedule in their blocks.  

To obtain a closed-form expression for the received reward, two parameters are added to the parameters of the Ethereum base model (Section \ref{ss:model}), namely the conflict rate and the number of processors. Note that we still assume that all blocks are valid, as in the base model.  The conflict rate $c$ is the percentage of conflicting transactions in a block. For example, $c$= 0.4 means that 40\% of the block's transactions are in conflict with other transactions in the same block. We note that according to \cite{dickerson2017adding}, the number of conflicting transactions in real blockchains is not very high since there are thousands of different contracts. The number of concurrent processors $p$ is the number of machines the miner has available in parallel. With $p$ processors and $c$ conflict rate, the slow down of performing the parallel verification process ($\delta$) is:

\begin{equation}
    \delta = (1 - \alpha V) * T\textsubscript{v} * (c + \frac{1 - c}{p})
    \label{eq:delta2}
\end{equation}

The fraction of blocks and rewards for verifying and non-verifying miners is based on the same equations as in Section \ref{ss:model}.

Apply the parallel verification to the previous example, with $c=0.4$ and $p=4$. Then, the slow down of performing the parallel verification as $\delta = 0.1749$. The fraction of blocks generated by the nine verifying miners is reduced from 0.9 to 0.888. Thus, the non-verifying miner would gain 0.012 more blocks and rewards ($\approx$ 12\% more than its invested $\alpha$). In other words, the fraction of blocks generated by the non-verifying miner increases from 0.1 to 0.112.

\subsection{Mitigation 2: Intentional Invalid Blocks}
\label{ss:risk_model}


In this section, we introduce a solution whereby Ethereum could allocate a special node for intentionally generating invalid blocks as a way to punish non-verifying miners. This special node is assigned a particular hash power (e.g., $\alpha$ = 0.04) of the total network hash power. The hash power of the special node simply represents the fraction of the invalid blocks to be purposely generated in the network. We assume this node to verify all blocks generated by other miners, and thus, it always works on the valid branch of the blockchain. The rationale behind this approach is that miners benefit from not verifying because all (or almost all) blocks are valid anyway. However, if incoming blocks could be invalid, the non-verifying miner could end up working on new blocks on top of the invalid ones. Consequently, the non-verifying miner would lose the rewards for those new blocks since other verifying miners will reject the blocks because they were built on top of invalid ones. Since this scenario includes non-valid blocks, we have no closed-form insights for this scenario. However, we will extensively study the result of injecting invalid blocks in the simulation results in Section \ref{s:results}. 

\section{Data Collection and Distribution Fitting}
\label{s:bench}

To obtain insights into the Verifier's Dilemma in Ethereum, we will parameterize the simulation with data collected from the Ethereum blockchain as well as data obtained from an experimental measurement system to measure the CPU time required for smart contract transactions. We managed to collect data and measure CPU time for about 324,000 contract-related transactions. To be of use to the simulation in Section \ref{s:evaluation}, we fit probability distributions to the transaction attributes (Gas Limit, Used Gas, Gas Price, and CPU time).

\subsection{Design of Data Collection Approach}
\label{ss:bench_data}

As we mentioned in Section \ref{ss:eth}, smart contracts are both created and executed through a transaction. In this section, we propose an automated data collection approach to collect the details (e.g., Gas Limit, Used Gas, Gas Price, and input data) of contract transactions (both contract-creation and contract-execution). For a contract-execution transaction, our approach also collects the details of the transaction that created the contract. Our approach makes use of the APIs provided by Etherscan\footnote{Etherscan is a block explorer platform for Ethereum that provides APIs with various functions related to accounts, contracts, blocks, transactions, etc.} to retrieve the details of transactions, and it is implemented as a Python script.

To get the CPU time data for contract-related transactions, we propose a system that is capable of measuring the CPU usage for smart contracts transactions by isolating the execution of the transactions from other computation and overhead (e.g., transaction validation and the PoW overhead). Our measurement system consists of two phases, namely, preparation and execution. During the preparation phase, we configure the blockchain and set up the Ethereum's global state. In addition, we initialize a set of Ethereum accounts to submit and execute transactions. During the execution phase, we construct, send, and execute transactions. We construct a transaction by setting its details or fields using the details for transactions we collected from Ethereum. Then, we use the accounts we initialised in the preparation phase to submit and execute the constructed transaction. The execution of a contract transaction requires three tasks, which are checking the validity of the transaction, running the data of the transaction on the EVM and finally updating the state upon successful execution. We place a timer before and after the execution of the transaction on the EVM. Once the transaction has been successfully executed, we record its Used Gas and the time it takes to run on the EVM. 

\begin{figure}
    \centering
    \includegraphics[scale=0.50]{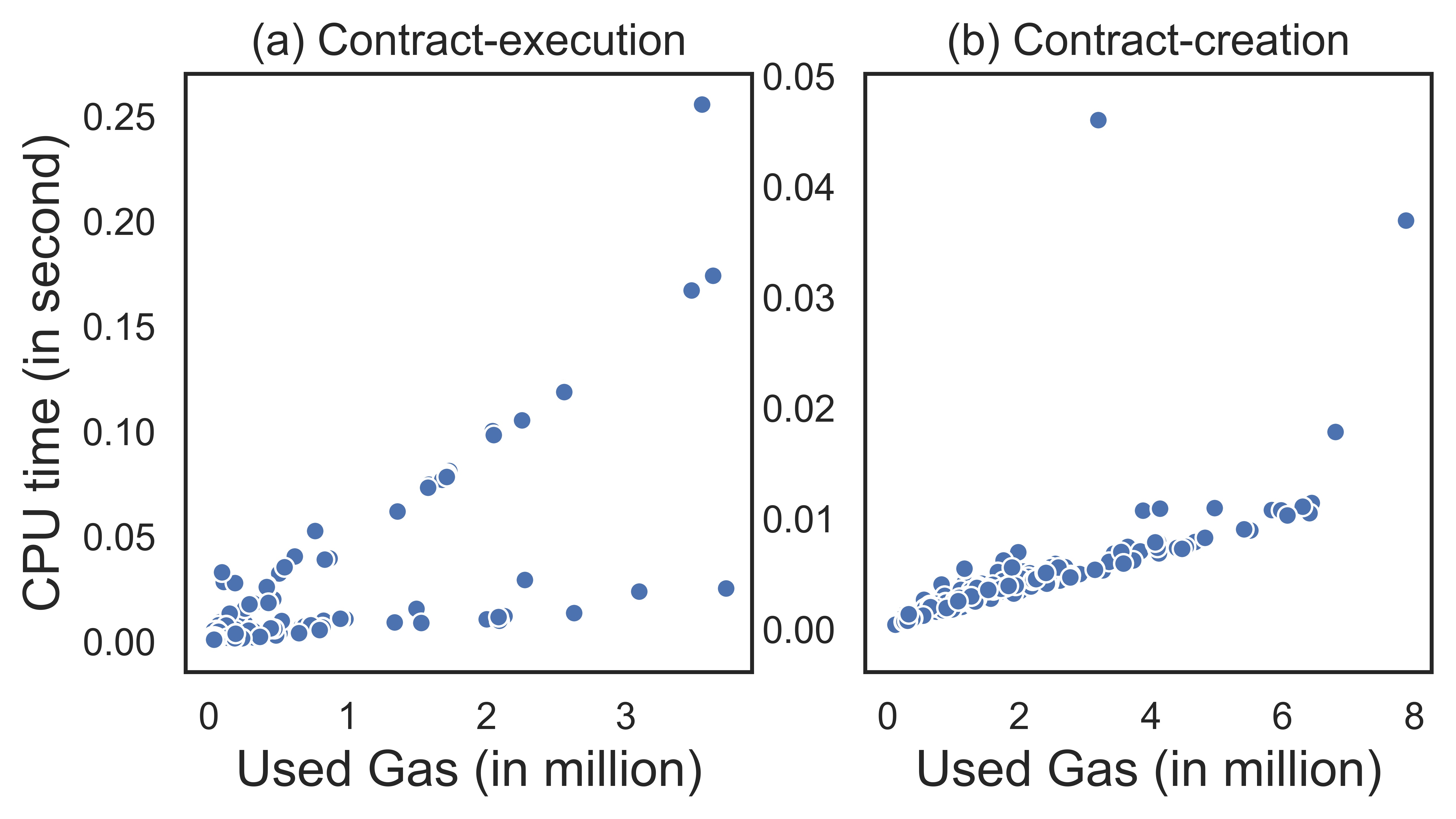}
  \caption{CPU Time (in second) versus Used Gas (in million) for (a) Contract-execution and (b) Contract-creation.}
    \label{bench_scatters}
\end{figure}

We managed to collect the details (including measuring the CPU time) of about 324,000 transactions (3,915 contract-creation and 320,109 contract-execution transactions). We selected those transactions randomly. Figure \ref{bench_scatters} shows the Used Gas in million versus the CPU Time in second for (a) contract-execution  and (b) contract-creation transactions. From Figure \ref{bench_scatters}, it is clear that the CPU usage is not proportional or linear with the amount of Used Gas, especially for contract-execution transactions.

The CPU time measured here is  from a single machine using the Python PyEthApp\cite{pyth} client. The machine is a desktop PC with a 3.40GHz Intel i7 CPU with 8GB RAM running on Windows 10. Each transaction is executed 200 times, and the average time is then calculated. The 95\% confidence interval is not reported here, but it is within 2\% of the average value.

\subsection{Fitting Distribution to Data}
\label{ss:Fitting}

To utilize the collected data in our simulation, we need to fit distributions with respect to attributes such as Used Gas, Gas Limit, Gas Price and CPU Time. To do this, we need to understand if there is dependency across these attributes. To understand the relationship between multiple variables or attributes in a data set we applied two different correlation methods, namely, Pearson \cite{pearson1900mathematical} and Spearman \cite{spearman1987proof}. 

The Pearson method assesses the linear relationship between variables, while the Spearman method assesses the monotonic relationship. In a linear relationship, the variables tend to change together at a constant rate, while in a monotonic relationship they tend to change together, but not necessarily at the same rate. Based on the correlation analysis results, we come to the following conclusions. (1) The CPU Time attribute has a strong positive non-linear correlation with Used Gas. (2) Gas Limit has a weak to a medium positive correlation with Used Gas. The correlation might come from the fact that the Gas Limit is always greater than or equal to the Used Gas. (3) Gas Limit has a weak to a medium positive correlation with the CPU Time. This correlation is slightly stronger for the creation set compared to the execution set. (4) Gas Price is independent of all other attributes, and, indeed, it does not have any relationship with different attributes.

We apply our fitting approach to the creation and execution sets separately. We fit a general probabilistic distribution to the Gas Price and the Used Gas values. We eventually decided to select Gaussian Mixture Models (GMMs) to fit the log of the data since none of the simple structured distributions fits the data particularly well. However, when considering the log of the data its shape resembles a normal distribution or a mixture of normal distributions.  

For the Gas Limit, it is appropriate to fit a uniform distribution, where the minimum value is the Used Gas and the maximum value is the block limit. This is because the Gas Limit is specified by the submitter of the transaction and it can take any value up to the block limit. Thus, the Gas Limit values will be drawn from a uniform distribution as follows:
\begin{equation}
Gas\ Limit \sim Unif(Used\ Gas, block\ limit)
\end{equation}
The current block limit is about $8*10^{6}$ unit of gas. For the CPU time, we fit a non-linear regression model (e.g., Random Forest Regressor) to predict the CPU time value from the given Used Gas value. 

For the CPU Time, we decided to select Random Forest Regression (RFR) \cite{breiman2001random} to train a model to be able to predict the CPU time value, given the Used Gas value. The reasons for selecting RFR is as follows. Firstly, it uses ensemble methods where different models (as opposed to a single model) are constructed in order to improve the accuracy of predictions. In addition, it is known to be robust against over-fitting, even when the number of decision trees increases \cite{breiman2001random}. Furthermore, RFR does not require any knowledge or assumptions about the distribution of the data \cite{yuchi2019evaluation}.

\begin{algorithm}
\footnotesize
\caption{The fitting and sampling procedure}
\begin{algorithmic}[1]

\Procedure{Fit a GMM to log}{$Gas\ Price$}       
    \State Determine $K$                        \Comment{Use AIC/BIC}
    \State Estimate $\sum_{i=1}^{K}\mu_i$, $\sum_{i=1}^{K}\sigma_i^2$,
    $\sum_{i=1}^{K}\phi_i$     \Comment{Use EM algorithm}
    \State $P$= GMM(K,$\sum_{i=1}^{K}\mu_i$, $\sum_{i=1}^{K}\sigma_i^2$,
    $\sum_{i=1}^{K}\phi_i$).fit(log($Gas\ Price$))
\EndProcedure
\Procedure{Fit a GMM to log}{$Used\ Gas$}       
    \State Determine $K$                        \Comment{Use AIC/BIC}
    \State Estimate $\sum_{i=1}^{K}\mu_i$, $\sum_{i=1}^{K}\sigma_i^2$,
    $\sum_{i=1}^{K}\phi_i$     \Comment{Use EM algorithm}
    \State $U$= GMM(K,$\sum_{i=1}^{K}\mu_i$, $\sum_{i=1}^{K}\sigma_i^2$,
    $\sum_{i=1}^{K}\phi_i$).fit(log($Used\ Gas$))
\EndProcedure
\Procedure{Fit a RFR to}{$Used\ Gas, CPU\ Time$}       
    \State Determine and optimise $d, s$                        \Comment{Use Grid Search CV}
    \State $T$= RFR($d,s$).fit($Used\ Gas, CPU\ Time$)
\EndProcedure
\Procedure{Sample attributes}{$S_P,S_U,S_L,S_T$}       
    \State $S_P$= $\exp(P.sample(n))$                       \Comment{Sample Gas Price}
    \State $S_U= \exp(U.sample(n))$                       \Comment{Sample Used Gas}
    \State $S_L= Unif(low= s_u, high= 8*10^{6}, size=n)$  \Comment{Sample Gas Limit}
    \State $S_T= T.predict(S_U)$    \Comment{Sample CPU Time}
\EndProcedure

\end{algorithmic}
\label{sim_algo}
\end{algorithm}

The procedure for fitting distributions to the attributes as well as sampling from such distributions is summarized in Algorithm \ref{sim_algo}. To fit a GMM to the log Used Gas and to the log Gas Price (line 1-8), we have to estimate the number of Gaussian components K, the mean $\mu_i$, the variance term $\sigma_i^2$ and the weight $\phi_i$ of each component. To determine $K$, we use Akaike Information Criterion (AIC) \cite{AIC} and Bayesian Information Criterion (BIC) \cite{BIC}. We tested $K$ values ranging from 1 to 100 and then selected the best $K$ according to these criteria. To determine the parameters for each component, we use the Expectation-Maximisation (EM) algorithm \cite{dempster1977maximum}. After estimating these parameters, we fit the GMM to the data. 

To fit a RFR model to learn and predict the CPU Time from a given Used Gas value (line 9-11), we have to determine and optimise the number of trees $d$ and the number of splits in each tree $s$. To optimise $d$ and $s$, we use a grid search technique with K-folds cross-validation (CV), where K = 10 as suggested by \cite{kohavi1995study}. We searched a number of values ranging from 10 to 500 and a number of values ranging from 1 to 300 to optimise $d$ and $s$, respectively. Then, we selected the best-tuned values for these parameters to fit the RFR model. After fitting the distributions to the attributes, we sample values for the transactions' attributes from the fitted distributions (line 12-16).

We implemented the algorithmic procedure in Python. We used the machine learning library \texttt{Scikit-learn} and utilised different packages from this library, namely GaussianMixture, RandomForestRegressor, GridSearchCV, and KFold. The GaussianMixture package contains the EM algorithm and the AIC/BIC criteria. We implemented a Python class to fit distributions to the attributes in the creation and the execution sets, respectively. In addition, we implemented a sampling method that takes as input the number of data points (transactions) to be simulated and returns the values of the simulated attributes as a tuple.  

The Appendix contains results of the fitting approach. In the main body of the paper we concentrate on aspects specific to the Verifier's Dilemma. 

\section{Simulator and Validation of Closed-Form Expressions}
\label{s:evaluation}

At the core of our model-based approach to analysing the Verifier's Dilemma is the publicly available \BlockSim simulator, which we extended to be able to study the Verifier's Dilemma. In this Section we report in Section \ref{ss:blocksim} on how we extended \BlockSim and in Section \ref{ss:validation} on the use of the simulator to validate the closed-form solutions derived in Section \ref{s:dilemma} and \ref{s:mitigation}. 

\subsection{\BlockSim Simulator Extension}
\label{ss:blocksim}

\BlockSim \cite{alharby2019blocksim,alharbyblocksim
} is a generic and extensible discrete-event simulation framework that can be used to analyse a variety of non-functional properties of various blockchain architectures. The \BlockSim simulator has been validated against data from real blockchain implementations as well as against measurement studies from the literature. The source code of the simulator is freely available. \BlockSim crosses three different layers (consensus, incentives, and network layers) and is based on core functional abstractions that are common across blockchains, such as blocks, transactions, fork resolution, and incentive distribution. The used implementation of \BlockSim includes the PoW-based consensus as used in both Bitcoin and Ethereum, and can be further extended and changed as required.  

To support the analysis of both the Ethereum base model and the solutions of parallel verification and intentional production of invalid blocks, we introduced the following modifications:

\textbf{The attributes of transactions:} We extended the Transaction class to include several attributes required by the model, which are Gas Limit, Used Gas, Gas Price, and CPU Time. Thus, each transaction created in our simulations has these attributes. 

\textbf{The distribution fitting class (\texttt{DistFit}):} We defined a new class named DistFit to fit probability distributions to the transaction attributes. This follows the procedure introduced in Section \ref{ss:Fitting}). We execute the distribution fitting once. During the simulation, when creating new transactions, we sample random values for these attributes from the fitted distributions.

\textbf{The number of processors (\textit{p}):} This dictates the number of processors a miner could use to verify transactions in parallel. To add this feature to the simulator, we extended the Node class by adding a new attribute named \textit{processors}. 

\textbf{The rate of conflicting transactions (\textit{c}):} This dictates the fraction of transactions that depend on other transactions in the system. To add this feature to the simulator, we introduced a new input parameter called \textit{conflict\_rate}. We also extended the Transaction class by adding a new attribute named \textit{dependency} for transactions, to distinguish between conflicting and non-conflicting transactions. Each transaction created will be assigned to a random value (True or False) for the \textit{dependency} attribute based on the \textit{conflict\_rate} parameter.  

\textbf{Parallel verification of transactions:} To add this feature to the simulator, we modified the execution of the block receiving event as follows. Upon receiving a new block, we distribute non-conflicting transactions between the different processors, after which the conflicting ones will be executed sequentially on a single processor. Hence, we count the time required to verify transactions in parallel by checking the CPU time attribute for transactions. Prior to starting the verification, the time for all processors is set to 0 (all processors are idle). During the verification process, we keep recording the time when each processor finishes the transaction at hand and pass a new transaction to start afterward.  

\textbf{The intentional production of invalid blocks:} To add this feature to the simulator, we first extended the Block class by adding a new attribute named \textit{validity} for blocks, to distinguish between valid and invalid blocks. Each block created will be assigned to a value (True or False) for the \textit{validity} attribute. Then, we set one of the miners to be the network node that always generates invalid blocks. The hash power of this node can be changed to reflect the fraction of invalid blocks to be generated in the network.

\subsection{Validation of Closed-Form Expressions}
\label{ss:validation}
To validate the closed-form solutions from Sections \ref{ss:model} and \ref{ss:par_model}, we need first to estimate the average time it takes to verify a block's transactions. The verification time depends on which transactions are included in the block. In particular, different transactions take different time as well as blocks may have very different number of transactions depending on the gas used by these transactions. Hence, we utilised the simulator to simulate different configurations of block limits (the limit is expressed in million (M) units of gas). For each configuration, we simulated 10000 blocks and the statistical results related to the block's verification time are given in Table \ref{tvb}. The table gives the minimum (min), the maximum (max), the mean, the median, and the standard deviation (SD) for the block's verification time, all in seconds.

\begin{table}[h]
\normalsize
\centering
\caption{The statistical results for the block's verification time ($T\textsubscript{v}$) in seconds for different block limits.}
\begin{tabular}{ c | c | c | c |c | c}
\hline
    & \multicolumn{5}{c} {Block's verification time ($T\textsubscript{v}$)} \\ \hline
    block limit & min & max & mean & median & SD \\ \hline
    8M & 0.03	&0.35	&0.23 & 0.24	&0.04\\ \hline
    16M & 0.16	&0.65	&0.46 & 0.47	&0.06\\ \hline
    32M & 0.51	&1.09	&0.87 & 0.87	&0.06\\ \hline
    64M & 1.06	&2.08	&1.56 & 1.56	&0.19 \\ \hline
    128M & 2.5	&3.75	&3.18 & 3.19	&0.19\\ \hline

\end{tabular}
\label{tvb}
\end{table}

To validate the closed-form expressions (Equations (\ref{eq:delta}) to (\ref{eq:delta2})) for both the Ethereum base model and the parallel verification solution, we compare the simulation results with that of the equations. We configured the simulator as follows. We set the block interval time to be 12.42 seconds, which is the minimum observed interval between blocks according to Etherscan\footnote{https://etherscan.io/chart/blocktime}.
The attributes (Gas Limit, Used Gas, Gas Price, and CPU Time) for transactions are generated from distributions, as discussed in Section\ref{ss:Fitting}. We set the number of miners to 10, where each miner controls 10\% of the total network hash power. Nine miners follow the protocol honestly by executing the verification process upon receiving a newly generated block, apart from one miner who skips the verification process. For the parallel verification, we set the number of processors to 4 and the conflict rate of transactions to 0.4. Then, we record the fraction of fee each miner receives at the end of the simulation. 

\begin{figure}
    \centering
    \includegraphics[scale=0.55]{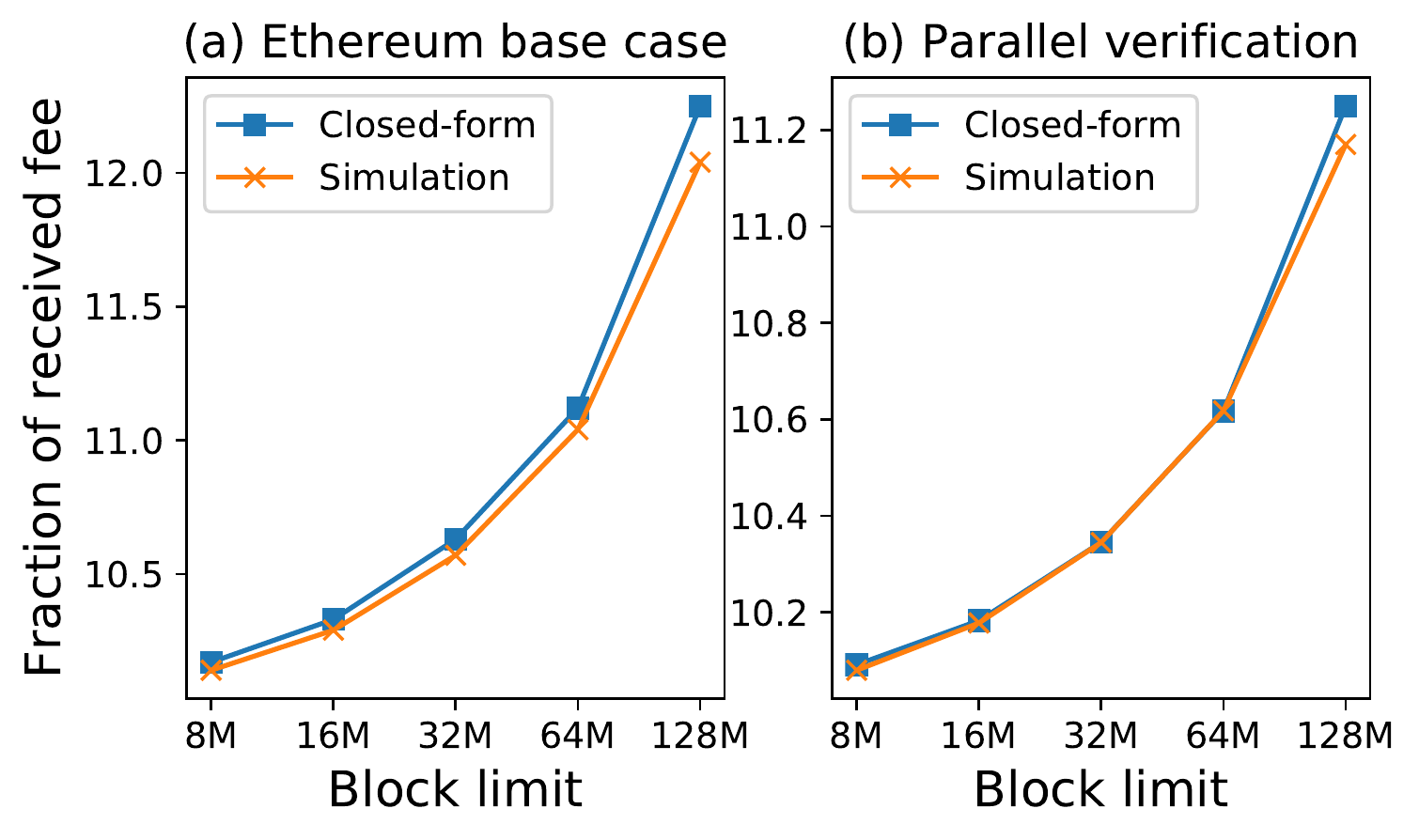}
  \caption{Results from the closed-form expressions and the simulation in the fraction of fee received by a non-verifying miner who has 10\% of hashing power, for (a) the Ethereum base model and (b) the parallel verification solution.}
    \label{validate_results}
\end{figure}

We run simulation experiments with different configurations of block limits (ranging from 8M to 128M). For each configuration, we simulated the equivalent of 3 days of running time of the Ethereum network and repeated this to have 100 independent runs. Figure \ref{validate_results} shows the validation of both the Ethereum base model and the parallel verification by presenting the results from the closed-form solutions as well as from the simulation.  The vertical axes shows the percentage of the received fee the non-verifying miner receives. One sees from Figure \ref{validate_results} that the non-verifying miner always wins, since in this scenario all blocks are valid, so the miner is never penalized for non verifying.  The gain can be a full percentage point or more as the block limit increases.   Various additional results will be discussed in Section \ref{s:results}.

We note that the simulation results slightly differ from that of the closed-form for the larger block limits. The closed-form expressions slightly overestimate the gain miners get from not validating blocks, but the differences are small. Several elements are modeled in more detail in the simulation than in the closed-form expressions, and these may contribute to randomness that causes a difference between closed-form and simulation. We believe that it is fair to conclude from Figure \ref{validate_results} that the closed-form expressions are close to the simulation results. 




\section{Results}
\label{s:results}
In this section, we present the main findings from our analysis of the Verifier's Dilemma, under the Ethereum base model as well as under the proposed mitigations of parallel verification and intentional production of invalid blocks. Our main metric of interest is the fee gained or lost by non-verifying miners in various scenarios.  

We summarize the main findings that follow from our discussion upfront:
\begin{itemize}
    \item The smaller the hash power a miner controls, the more advantage the miner would gain from skipping the verification process.
\item In today's Ethereum, miners gain relatively little from skipping the verification (less than 2\% of the invested hash power).  This is because the block limit in Ethereum is currently small. 
\item In the future, the Ethereum block limit is expected to increase. In that case, skipping verification becomes considerably more lucrative.  This is under the assumption that most miners honestly verify and invalid transactions are rare.
    \item Parallel verification reduces the benefits miners would get from not verifying blocks. This is especially true if the conflict rate is small and the number of parallel processors is large.
    \item The mitigation approach to purposely introducing invalid blocks in the network can significantly reduce the benefits of non-verifying miners. This is especially true if the rate of invalid blocks is large or the block limit is small.  In this case, miners may be better off to verify. 
   
\end{itemize}

\subsection{Ethereum Base Model}
\label{ss:seq_results}
\begin{figure}
    \centering
    \includegraphics[scale=0.55]{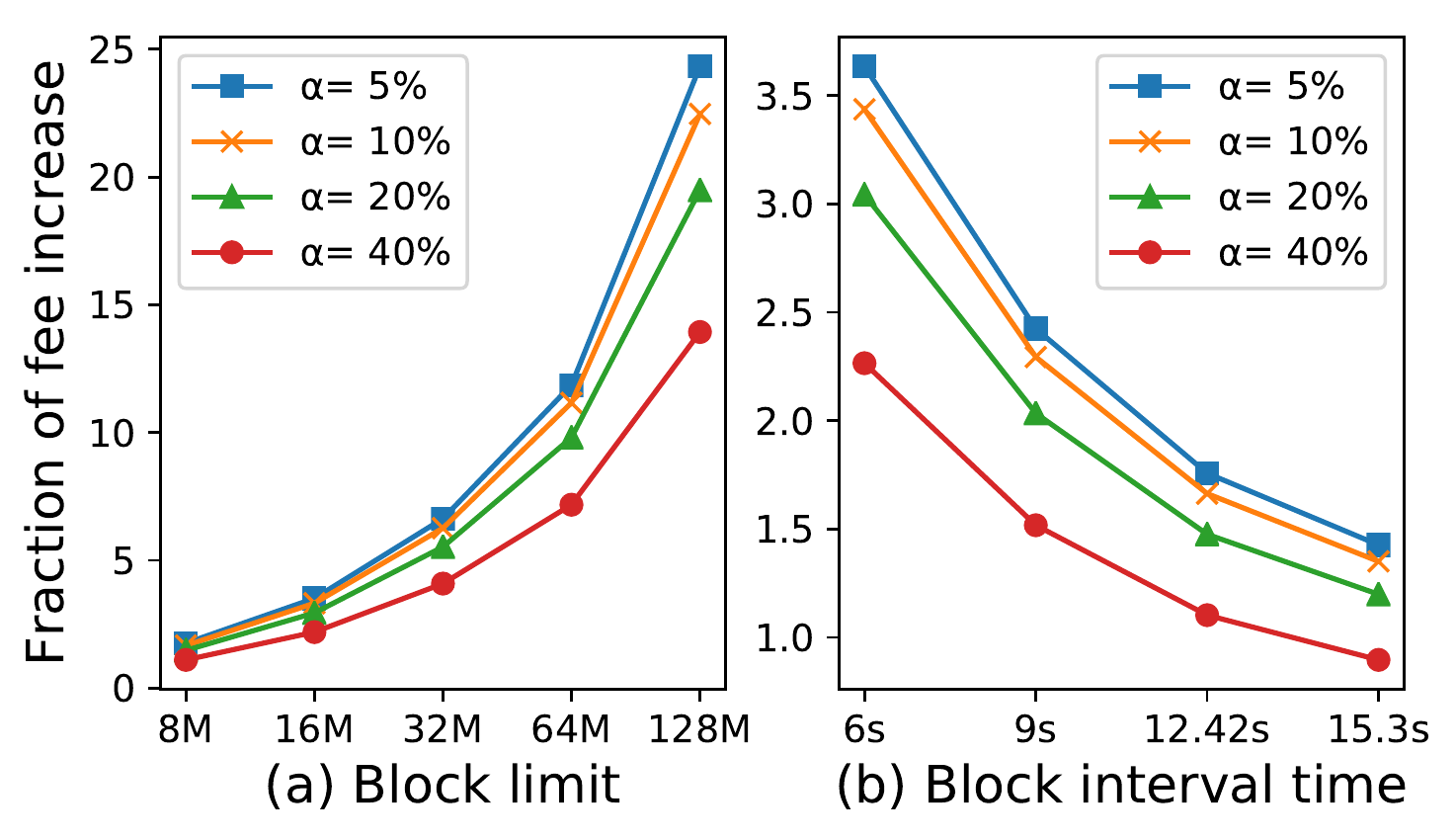}
  \caption{The percentage of fee increase for a non-verifying miner with the Ethereum base model: (a) different block limits and (b) different block interval time.}
    \label{seq_verification}
\end{figure}

Figure \ref{seq_verification} shows the percentage of fee increase a non-verifying miner would gain, for different block limits and different block interval times. The four curves in each of the two plots of Figure \ref{seq_verification} indicate different fractions of the total hash power owned by the non-verifying miner. In Figure \ref{seq_verification}(a), we consider a block interval time of 12.42 seconds. In Figure \ref{seq_verification}(b), we consider a block limit of 8M, which is the block limit currently used in Ethereum.


From Figure \ref{seq_verification} we conclude that for the current implementation of Ethereum (block limit = 8M and block interval time is between 12 and 15 seconds), the percentage fee increase is small (less than 2\% of the invested hash power). Yet, this percentage increases significantly with the block limit or the reduction of the block interval time. For instance, a non-verifying miner with $\alpha$ = 0.05 would increase its gain from 1.7\% for small blocks to a remarkable 22\% when the block limit is pushed from 8M to 128M. In addition, we can see that the smaller the hash power a miner controls the larger the increase the miner gets when not verifying blocks. For example, a miner with $\alpha$ = 0.05 can increase its fraction of fee to 24\% when the block limit is 128M, while it only increases its fraction to about 14\% if $\alpha$ = 0.40. This is because a miner has to verify all the blocks that were mined by others, which amounts to $(1 - \alpha)$ of the network blocks, as we discussed in Section \ref{ss:model}. In other words, in Ethereum, small miners spend more time on verification than large miners because they receive more new blocks from other miners.  Therefore, small miners have more to gain from stopping with verifying.  

\subsection{Parallel Verification}
\label{ss:par_results}

\begin{figure*}
    \centering
    \includegraphics[scale=0.55]{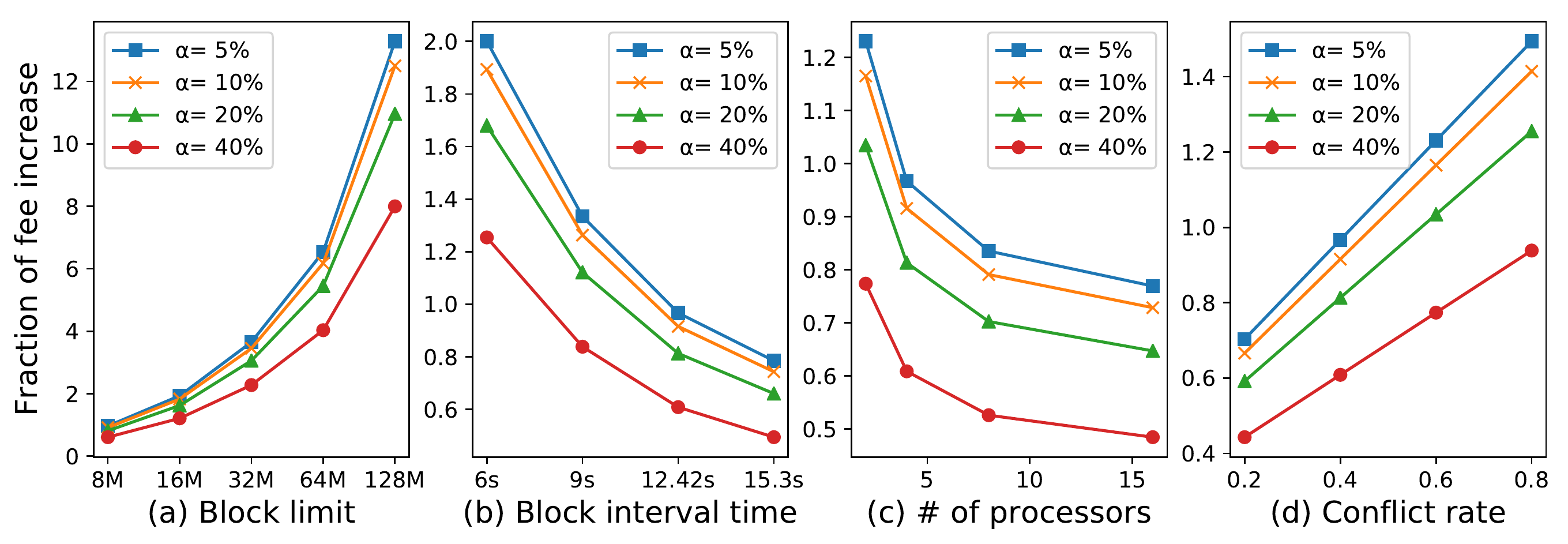}
  \caption{The percentage of fee increase for a non-verifying miner with parallel verification: (a) different block limits, (b) different block interval time, (c) different number of processors and (d) different conflict rates.}
    \label{par_verification}
\end{figure*}

Parallel verification of transactions is a solution that we proposed in Section \ref{ss:par_model} to minimize the advantage non-verifying miners would gain by reducing the overall time required for the verification process. Figure \ref{par_verification} shows the percentage of fee increase that a non-verifying miner would gain, for different block limits, different block interval times, different number of processors and different conflict rates for transactions. As in Section \ref{ss:seq_results}, the different curves represent different hash powers for the non-verifying miner. 

From Figure \ref{par_verification} we see that although the percentage of fee increase rises with the block limit or the reduction of the block interval time, the advantage is reduced almost to half that of the Ethereum base model (see Figure \ref{seq_verification}).  This is for modest parallelization, with only 4 processors and a conflict rate of 0.4. In addition, from (c) and (d) we see that the advantage decreases further with the increase of the number of processors or with a small rate of conflicting transactions. For instance, assume an 8M block limit and 0.4 conflict rate.  Then the increase a non-verifying miner with $\alpha$= 0.10 would get goes down from about 1.2\% to 0.7\%, when increasing the number of processors from 2 to 16. To summarize, the  advantage  a  miner  would  gain  by  skipping  the  verification is  minimized  when  shifting  from  the Ethereum base model to the parallel  verification solution. The degree of reduction depends on the conflict rate and the number of concurrent processors.

  

\subsection{Production of Invalid Blocks}
\label{ss:invalid_results}

\begin{figure}
    \centering
    \includegraphics[scale=0.55]{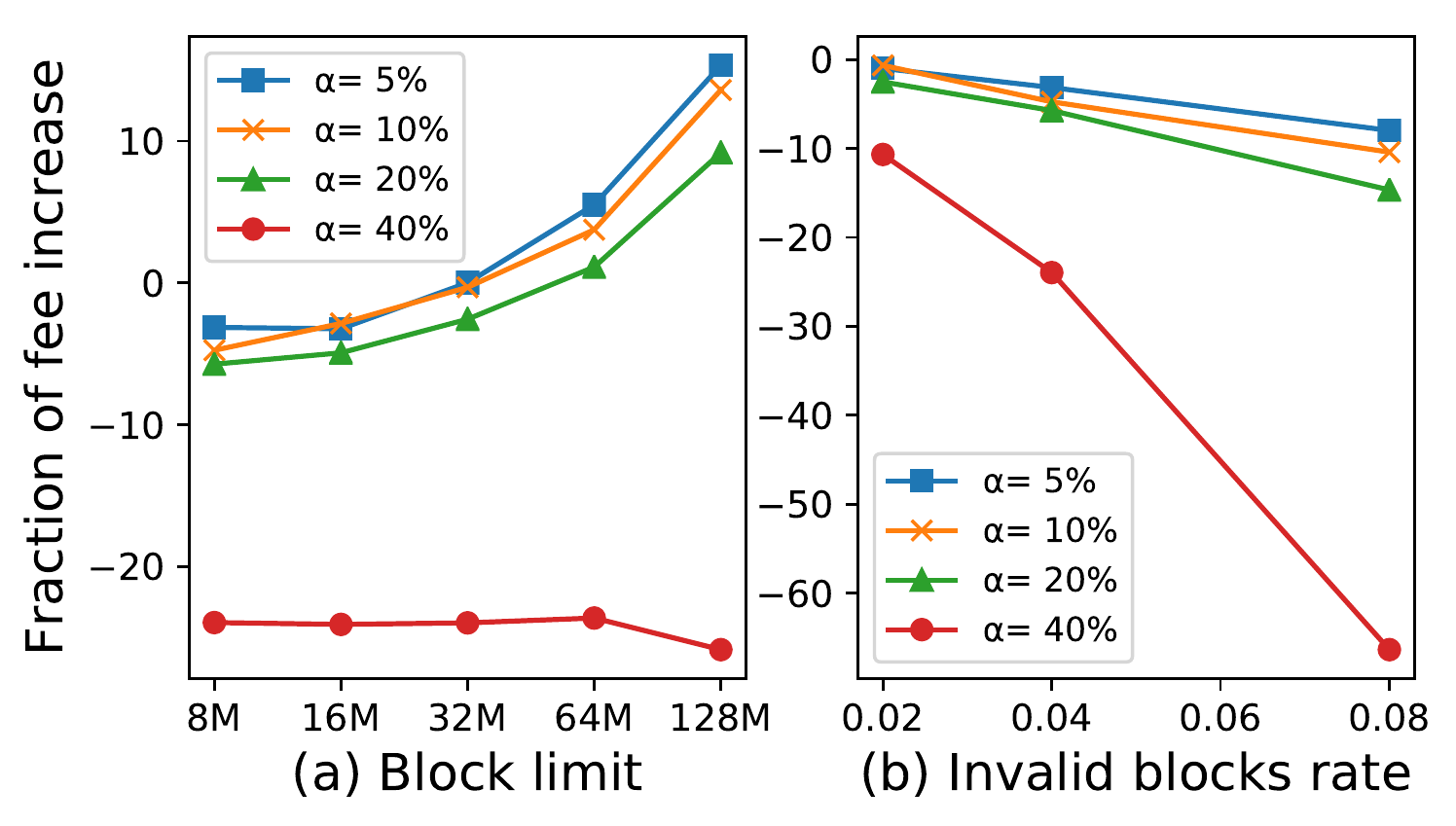}
  \caption{The percentage of fee increase for a non-verifying miner with the intentional production of invalid blocks: (a) different block limits and (b) different rates of invalid blocks.}
    \label{invalid_blocks}
\end{figure}



The idea behind intentionally introducing invalid blocks (Section \ref{ss:risk_model}) is to punish non-verifying miners. To assess if this approach can be useful, we modify the \BlockSim simulation classes to account for the possibility of having invalid blocks. We run simulation experiments with different configurations of block limits and different rates of invalid blocks. The rate of invalid blocks refers to the hash power of the special node that only generates invalid blocks. For each configuration, we simulated 1 day of the Ethereum network and reported the average results obtained from 100 independent runs. For these experiments, we considered a block interval time of 12.42 seconds.

Figure \ref{invalid_blocks} shows the percentage of fee increase that a non-verifying miner would gain given some fraction of invalid blocks in the network, for different block limits and different rates of invalid blocks. 
The different curves represent different hash powers for the non-verifying miner.
For Figure \ref{invalid_blocks}(a), we consider a block interval time of 12.42 seconds and an invalid blocks rate of 0.04. For Figure \ref{invalid_blocks}(b), we consider a block interval time of 12.42 seconds and a block limit of 8M.

From Figure \ref{invalid_blocks} we see that the fee increase for non-verifying miners is significantly reduced when inserting invalid blocks in the network. For instance, the fee increase a non-verifying miner with $\alpha$= 0.10 would get decreases from about 22\% to 13.6\%, when the rate of invalid blocks is 0.04 and the block limit is 128M. 

Even more interesting, non-verifying miners might get {\em less} reward than one would expect based on their hash power.  That is, we establish a situation in which verifying is preferred over not verifying.  This is especially pronounced when the block limit is small or when the rate of invalid blocks is large.  For example, a non-verifying miner with $\alpha$= 0.10 would lose about 5\% fee when the block limit is 8M and the rate of invalid blocks is 0.04. That means that conducting the verification process in that case is more profitable than skipping it.

We also note that miners with large hash powers (e.g., $\alpha \geq$ 0.20) are affected more when not verifying blocks, compared to miners with small hash powers. For example, a non-verifying miner with $\alpha$ = 0.05 can lose about 3\% of its expected fraction of fee when the block limit is 8M and the rate of invalid blocks is 0.04, while it would lose about 24\% of its expected fraction of fee if $\alpha$ = 0.40.

To summarize, purposely introducing invalid blocks into the blockchain could discourage miners from not verifying received blocks. The degree of deterrence depends on the rate of invalid blocks in addition to other blockchain configurations such as block limit.


\section {Discussion}
\label{s:discussion}

In Section \ref{s:results}, we showed the advantage a non-verifying miner may get from not verifying blocks in the Ethereum base model. In addition, we showed how parallel verification and intentional production of invalid blocks could mitigate against such behaviour, potentially even making skipping the verification worse than verifying itself. In this section, we discuss the possible threats to the validity of our evaluation.

\textbf{Execution time of transactions:} The CPU time to verify each transaction is based on the measurements obtained from a particular machine. In reality, miners might use different and possibly much more powerful machines and the specifications of machines are expected to improve in the future. The key indicator for the relative importance of the verification effort is how it compares to the effort spent on mining new blocks, which is determined predominantly by the PoW difficulty. However, even with more powerful machines and/or increased PoW difficulty, the Verifier's Dilemma will become a problem at some point if the block limit reaches a particular threshold or if more complex contracts are permitted.

\textbf{Different types of transactions:} We studied the Verifier's Dilemma assuming that all transactions in the network are contract-related. However, there are many financial transactions in Ethereum and since these can be verified very quickly the advantage of not verifying blocks may not be as large as in Section \ref{s:results}. In that sense, our analysis should be considered a worst case analysis. Either way, the main insights derived from our analysis remain valid, even if exact values and results may be different.  

\textbf{Full blocks of transactions:} We also assumed that all blocks are filled up with transactions, but in reality it is possible to have non-full or even empty blocks. In that case the advantage of not verifying blocks will also be less than in Section \ref{s:results}. However, by design the block reward is decreasing and is expected to be removed eventually \cite{8661611}. When there is no block rewards, miners will be much more encouraged to fill up their blocks with transactions to maximise their rewards.  Irrespective, our analysis can be considered a worst case analysis and the main insights derived from our analysis remain valid, even if exact values and results may be different. 

\textbf{Parallel verification of transactions:} Parallel verification discourages skipping verification. However, the implementation of parallel verification on a real blockchain system is an open research problem \cite{alharby2017blockchain,
yang2019empirically}. Complications arise because, first, a miner needs to attach a table to its block in order for the verifier to know which transactions can be run on parallel. Producing the table is not simple since the miner has to figure out conflicting and non-conflicting transactions \cite{dickerson2017adding}. A second issue is that one needs to trust the miners to produce the correct table. And, finally, the EVM should be updated to support multi-threading.

\textbf{Intentional insertion of invalid blocks:} This approach does not only make skipping the verification a less beneficial strategy, it often makes verifying the prefered strategy. Although this solution can be easily adopted in Ethereum, its introduction would likely face some challenges. Producing invalid blocks in the network decreases the performance of the system and it will impose an extra task for honestly verifying miners as they are expected to verify those invalid blocks and then reject them.  So, in practice, one would expect Ethereum to be very hesitant adding such overhead to the system.  

\textbf{Different consensus algorithms:} We studied the Verifier's Dilemma for the PoW protocol. However, Ethereum and other blockchains are planning to move to more efficient protocols such as Proof of Stake (PoS) \cite{wang2019survey,buterin2017casper}. Since alternatives to PoW can be expected to be computationally much more efficient than PoW, one would expect that the computation associated with verification becomes relatively more important. This would increase the pressure on miners to consider not to verify blocks. Clearly, studying the impact of the verification process under different consensus protocols is of interest. Within PoS, for instance, miners might be given a specific time window to finish and propose a block. If the miner spends a long time doing the verification process, it might not be able to finish the block on time, losing the rewards \cite{pontiverossluggish}.

\section{Related Work} 
\label{s:related}

The \textit{Verifier's Dilemma} was first identified by Luu et al. \cite{luu2015demystifying}, who showed that rational miners would be motivated to skip the verification process to gain an advantage in the race to mine the next blocks. Related to this idea is the mining strategy proposed in \cite{pontiverossluggish}, whereby a malicious miner purposely design smart contracts that are computationally expensive to verify, to slow down other miners. In response, \cite{luu2015demystifying,pontiverossluggish} showed the profitability of skipping the verification process in scenarios in which computationally intensive smart contracts were introduced. 

This work left unanswered several questions. In particular, it was not known if the above attack was practical, nor was it clear how different miners with different hash powers might benefit from not verifying blocks. In this paper, we address these limitations by evaluating the implications of the Verifer's Dilemma using real Ethereum smart contracts transactions. In addition, we assess both current and future settings of Ethereum, taking into consideration the hash power of miners.



Several solutions in the literature have been proposed to make the verification of complex transactions a more efficient task in order to avoid the Verifier's Dilemma. In \cite{luu2015demystifying}, the authors proposed a solution in which complex transactions are divided into various smaller transactions that can be incorporated in various blocks. 
In \cite{dickerson2017adding,yu2017smart,anjana2019efficient}, the authors proposed solutions for executing and verifying smart contracts in parallel. They showed that such solutions could speed up the execution/verification time of contracts compared to that of a sequential solution. 
In addition, several off-chain solutions (e.g., \textit{TrueBit} \cite{teutsch2017scalable}, \textit{YODA} \cite{das2018yoda} and \textit{Arbitrum} \cite{kalodner2018arbitrum}) for efficient computation of computationally expensive smart contracts have been proposed as an alternative to the protocol used in Ethereum. In these solutions, only a small set of nodes, instead of all nodes, has to perform the verification of complex contracts. Those nodes will be rewarded if they perform the verification correctly, or otherwise, a penalty will be imposed.

In \cite{dickerson2017adding,yu2017smart,anjana2019efficient}, however, the authors did not investigate the parallel verification of smart contracts as a mitigation solution to the Verifier's Dilemma. 
In this paper, we propose and evaluate the parallel verification as a solution to reduce the advantage miners would get from not verifying. Besides, we propose and evaluate the intentional production of invalid blocks as a new solution to punish non-verifying miners. We inspired the idea of injecting invalid blocks in the network from \cite{teutsch2017scalable}.

\section{Conclusion}
\label{s:conclusion}
This paper provides an extensive analysis of the Verifier's Dilemma, following a data-driven, model-based approach that combines closed-form expressions with discrete-event simulation and utilizes machine learning techniques to parameterize and configure probability distributions used by the simulator. This is the first extensive analysis of the Verifier's Dilemma we are aware of. The insights we gained in this paper can be of assistance in anticipating the implications of the Verifier's Dilemma under future developments, e.g., when the block limit increases in Ethereums, or when Proof of Work is replaced. Of particular importance for the fairness of blockchain systems is that our analysis shows that small miners are more impacted by the verification demands, and will be more tempted not to verify. Our results also indicate that, counter-intuitively, problems associated with the Verifier's Dilemma exacerbate if there are less invalid transactions.  This leads to the insight that future blockchain systems may operate better if designers or operators assure that some transactions are invalid. We suggest that similar analysis as reported in this paper should be carried out for future system designs and operational developments, to anticipate the consequences of the Verifier's Dilemma.   

\section{Appendix}
\label{s:append}

\subsection{Evaluation of the Fitting Algorithm}

To fit distributions to the transactions' attributes (see Section \ref{ss:Fitting}), we use two different models, namely, GMMs and RFR. In this section, we evaluate the performance of these models using two approaches. The first one is a visual evaluation by comparing the kernel density estimation (KDE) of the original data with that of the sampled data generated from the models. The second approach is to evaluate the performance of the models using some suitable metrics such as $r\textsuperscript{2}$, where possible.

\subsubsection{RFR Evaluation}
To assess the accuracy of the RFR models, we utilised different score metrics such as mean absolute error (MAE), root mean squared error (RMSE), and the coefficient of determination (R\textsuperscript{2}). We measured the accuracy of an RFR model for seen and unseen data to ensure the generality of the model and the robustness of the model against over-fitting. We measured the performance of the model by relying on K-folds cross-validation, where K = 10, as suggested by \cite{kohavi1995study}. We refer to the performance of the model with seen and unseen data as the \textit{training results} and the \textit{testing results}, respectively. 

\begin{table}[h!]
\center
\caption{The performance evaluation of the RFR regression models for both creation and execution sets. }
\begin{tabular}{ l |  l l  l | l l  l}
\hline
    & \multicolumn{3}{c}{Training Results} & \multicolumn{3}{c}{Testing Results} \\
    & MAE & RMSE & R\textsuperscript{2} & MAE & RMSE & R\textsuperscript{2} \\ \hline
    Creation Set & 34.29 & 355.12 & 0.96 &78.47 & 900.20 &0.82 \\ \hline
    Execution Set & 25.63 & 162.74 &0.99 & 29.39 & 426.59 &0.93 \\ \hline

\end{tabular}
\label{rfr}
\end{table}

Table \ref{rfr} shows different score metrics (MAE, RMSE, and R\textsuperscript{2}) to measure the performance of the RFR regression models used to predict the CPU Time values for both the creation and execution sets. From Table \ref{rfr}, we can see that RFR models perform well on both seen and unseen data for both creation and execution sets. This indicates the generality of the RFR models as well as the avoidance of over-fitting. 


\begin{figure}
    \centering
    \includegraphics[scale=0.54]{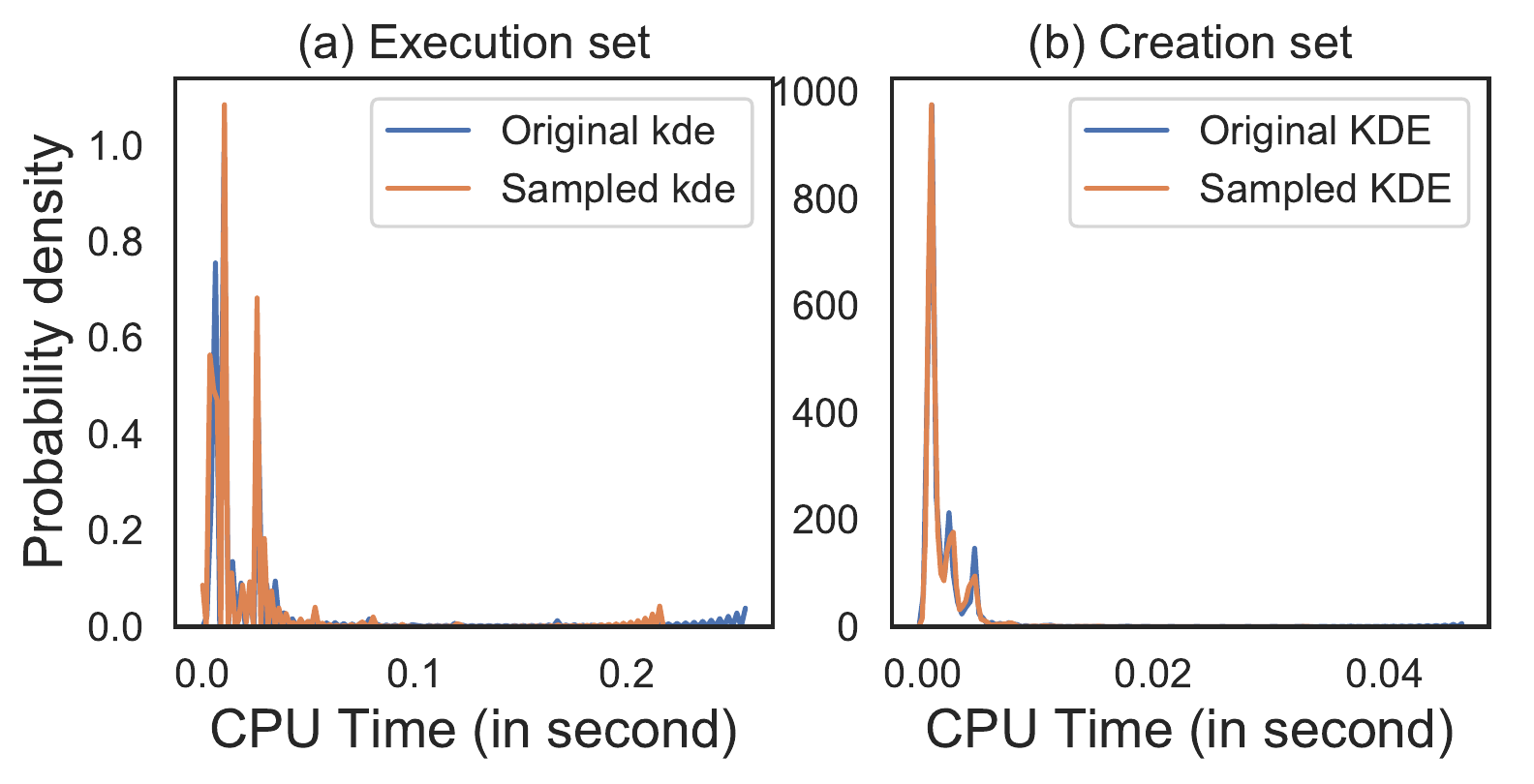}
  \caption{KDE for original and sampled CPU time for both execution set (left) and creation set (right).}
    \label{KDE_time}
\end{figure}

In addition to the evaluation metrics, we compared the KDE of the original CPU Time data with that of the sampled one. This is to assess if the samples generated from the RFR models are similar to the original data. From Figure \ref{KDE_time}, it is clear that the KDE for the sampled data looks very similar to that of the original one.

\subsubsection{GMMs Evaluation}
\begin{figure}[h!]
    \centering
    \includegraphics[scale=0.52]{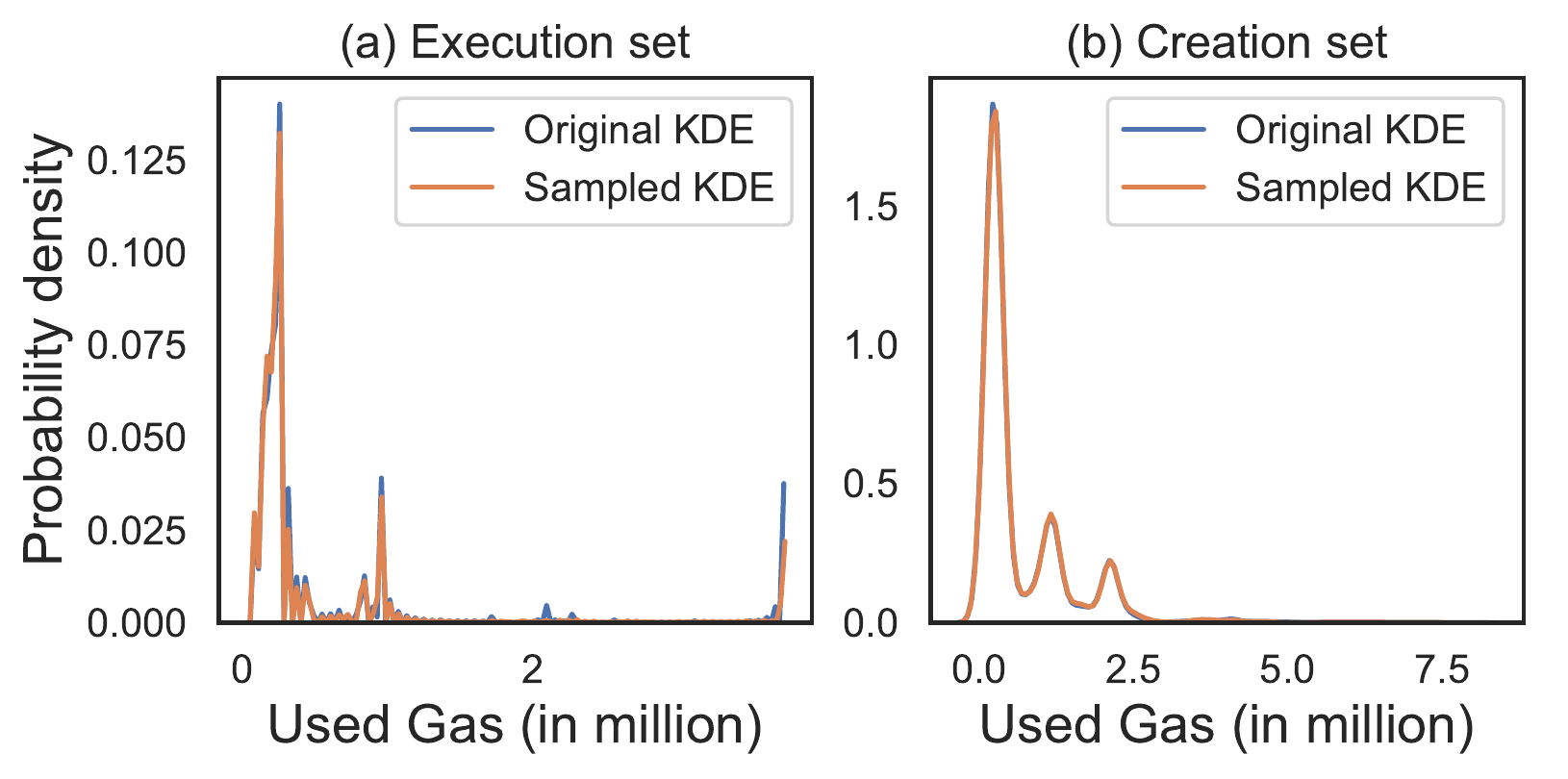}
  \caption{KDE for original and sampled Used Gas for both execution set (left) and creation set (right).}
    \label{KDE_used}
\end{figure}

\begin{figure}[h!]
\centering
\includegraphics[scale=0.52]{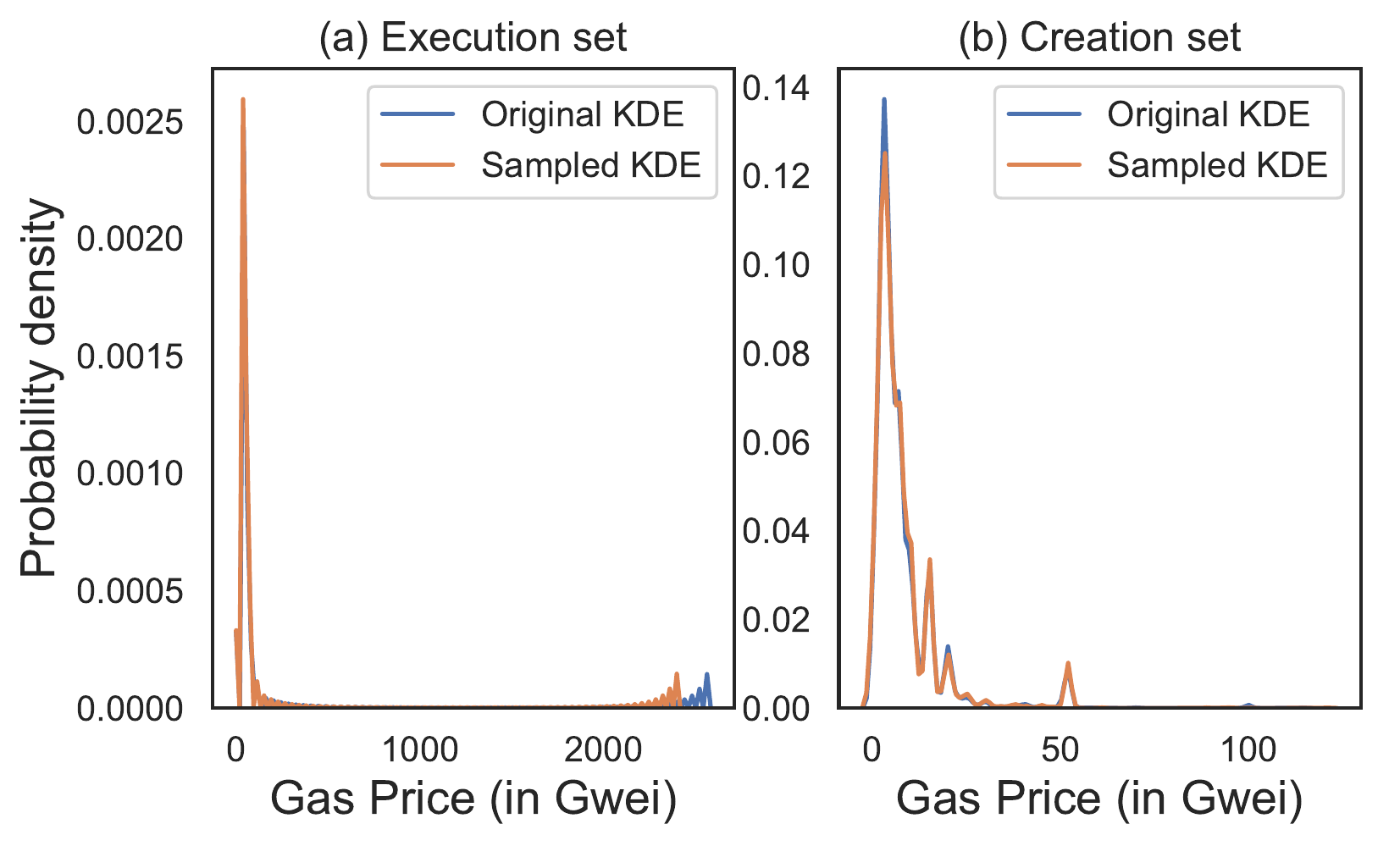}
\caption{KDE for original and sampled Gas Price for both execution set (left) and creation set (right).}
\label{KDE_price}
\end{figure}

It is not possible to validate 1D GMM models using performance metrics as we did with the RFR model. However, we compared the KDE for the original Used Gas data with that of the sampled one. This is to assess whether the samples generated from the GMMs models are similar to the original data. From Figure \ref{KDE_used}-\ref{KDE_price}, it is clear that the KDE for the sampled data looks very similar to that of the original one for both Used Gas and Gas Price. 

\bibliographystyle{IEEEtran}
\bibliography{references}

\end{document}